\documentclass[final,5p,times,twocolumn]{elsarticle}

\usepackage{amssymb} 
\usepackage{textcomp}
\usepackage{braket}
\usepackage{sidecap}
\usepackage{todonotes}
\usepackage{hyperref}
\usepackage{floatrow}
\usepackage{url}

\def\mathbi#1{\ensuremath{\textbf{\em #1}}}
\def\Q{\ensuremath{\mathbi{Q}}}

\journal{Journal of Magnetism and Magnetic Materials}
\begin{document}
\begin{frontmatter}
\title{}

\title{Insights into the High Temperature Superconducting Cuprates from Resonant Inelastic X-ray Scattering}
\author{M. P. M. Dean}
\ead{mdean@bnl.gov}
\address{Department of Condensed Matter Physics and Materials Science, Brookhaven National Laboratory, Upton, New York 11973, USA}

\begin{abstract}
Recent improvements in instrumentation have established resonant inelastic x-ray scattering (RIXS) as a valuable new probe of the magnetic excitations in the cuprates. This article introduces RIXS, focusing on the Cu $L_3$ resonance, and reviews recent experiments using this technique. These are discussed in light of other experimental probes such as inelastic neutron scattering and Raman scattering. The success of these studies has motivated the development of several new RIXS spectrometers at synchrotrons around the world that promise, among other improvements, 5-10 times better energy resolution. We finish by outlining several key areas which hold promise for further important discoveries in this emerging field.
\end{abstract}
\end{frontmatter}

\section{Introduction\label{Sec:Intro}}
Understanding high temperature superconductivity (HTS) in the cuprates has been one of the defining problems of condensed matter physics for the last quarter of a century \cite{Norman2011}.  At the core of the problem is the quest to characterize the nature of the ground state and the low energy excitations that define the normal state from which HTS emerges.  This has driven the development of several spectroscopic techniques including angle-resolved photoemission (ARPES) \cite{Yoshida2012} and scanning tunneling spectroscopy (STS) \cite{Fujita2012STS} as probes of electronic structure, and inelastic neutron scattering as a probe of magnetism \cite{Tranquada2014}. Indeed, these techniques have provided numerous important insights into the physics of the cuprates, including the emergence of Fermi arc features in the electronic spectral function \cite{Fujita2012STS} and the resonance phenomenon in which the dynamical magnetic  susceptibility changes through the superconducting transition \cite{Tranquada2014}. In recent years, instrumentation for resonant inelastic x-ray scattering techniques (RIXS) with both soft and hard x-rays has also improved dramatically \cite{Ament2011}, allowing this technique to directly measure magnetic excitations in several materials such as the cuprates \cite{Hill2008, Braicovich2009, Braicovich2010}, nickelates \cite{Ghiringhelli2009}, pnictides \cite{Zhou2013} and iridates \cite{Kim2012, Kim2012_327, Yin2013}.

This Current Perspectives article describes recent experimental progress in soft x-ray RIXS studies of magnetic excitations in the cuprates with a particular focus on the doping dependence of the magnetic excitation spectrum and how this relates to superconductivity \cite{Braicovich2010, LeTacon2011, Dean2012, Ghiringhelli2012, DeanBSCCO2013, DeanLSCO2013, LeearXiv2013}. We start by outlining the RIXS technique and the basics of the cuprate phase diagram. This article then describes some of the insights gained into the cuprates; first in the underdoped and optimally doped cases, before moving on to the overdoped case, including both the measurements themselves and also their relationship with other experimental probes such as inelastic neutron scattering and Raman scattering. We end by discussing upcoming improvements in RIXS instrumentation and outline some opportunities for future experiments on stripe-ordered cuprates and heterostructures, as well as extending RIXS to probing phonons.

\begin{figure}
\centering
\includegraphics[width=\linewidth]{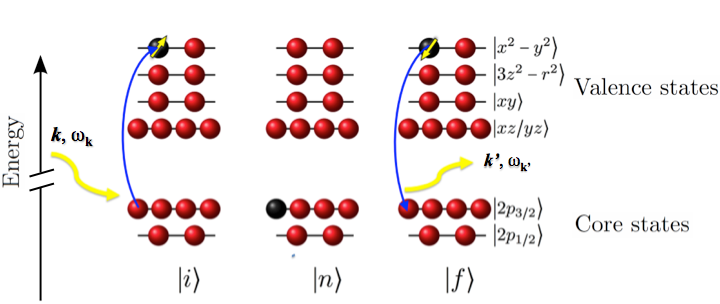} %
\caption{A schematic of the direct RIXS process showing the case of the $L_3$-edge resonance for a Cu $3d^9$ ion at 931~eV.  The initial $\ket{i}$, intermediate $\ket{n}$, and final $\ket{f}$, states are shown from left to right. Red spheres denote states filled by electrons and black spheres denote holes in up or down spin states. Incoming and outgoing photons are represented as wavy yellow lines and the blue arrows depict transitions. In this process spin excitations are created with energy, $\hbar \omega_{\mathbi{k}} - \hbar \omega_{\mathbi{k}^{\prime}}$, and momentum, $ \hbar \mathbi{q} = \hbar \mathbi{k} - \hbar \mathbi{k}^{\prime}$.
}
\label{Fig:RIXSprocess}
\end{figure}

\subsection{Resonant Inelastic X-ray Scattering\label{Sec:RIXS}}
\begin{figure}
\centering
\includegraphics{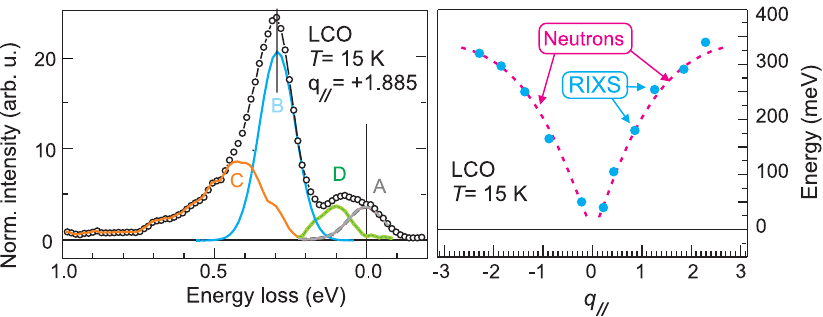} %
\caption{The first unambiguous measurement of a single magnon excitation by RIXS in La$_2$CuO$_4$ reproduced from Ref.~\cite{Braicovich2010}. Left: An example spectrum with the main single magnon excitation labeled as B. Right: A comparison between the peak in the RIXS spectrum (blue dots) \cite{Braicovich2010} and the magnetic dispersion determined by fitting neutron scattering results \cite{Coldea2001}. Copyright 2010 by The American Physical Society.}
\label{Fig:Braicovich}
\end{figure}

Resonant inelastic x-ray scattering, abbreviated RIXS, is an x-ray spectroscopic technique in which one measures the change in energy and momentum of x-rays that scatter from a material. This technique has great potential for probing the low energy excitations in correlated electron systems such as the cuprates, as it can be used to study electronic, magnetic, lattice and orbital excitations. Furthermore, it is element and orbital resolved, bulk sensitive and compatible with small samples \cite{Ament2011}.  Figure \ref{Fig:RIXSprocess} illustrates the RIXS process for the case that will form the focus of this article:  the $L_3$ resonance in the Cu $3d^9$ ion present in the cuprates. In the initial state, $\ket{i}$, there is one hole in the Cu $3d$ valence band, which is filled by the incident x-ray exciting a $2p_{3/2}$ core electron to form a highly energetic intermediate state, $\ket{n}$. Due to the strong spin-orbit coupling of the core hole, the orbital angular momentum of the photon can be exchanged with the spin angular momentum of the valence hole in order to create a spin flip excitation while conserving total (spin + orbital) angular momentum \cite{Ament2009}. The core hole is then filled to form the final state, $\ket{f}$, which contains a spin flip magnetic excitation distributed throughout the lattice. Such an experiment, in which the core electron is promoted into the valance band, and an electron from a \emph{different} state fills the core hole, is called \emph{direct} RIXS . This is distinct from indirect RIXS were an electron is excited into a high energy vacant state well above the chemical potential and an electron from the \emph{same} state fills the core hole \cite{Brink2006}. When direct RIXS is not forbidden it is the dominant process and it is this process that we focus on in this article \footnote{The reader is referred to Refs. \cite{Kotani2001, Ament2011} for a discussion of indirect RIXS.}. For many years, it was believed that single magnons could not be observed in $L$-edge RIXS studies of the cuprates without a simultaneous orbital excitation \cite{deGroot1998}. Only in 2009 was it demonstrated that single magnon excitations are only forbidden when the spin orientation is perpendicular to the CuO$_2$ planes \cite{Ament2009}. Given that the spins in magnetically ordered cuprates lie parallel to the CuO$_2$ planes, this implied that given sufficient energy resolution and signal-to-noise ratio, these excitations should be experimentally accessible. The energy resolution for soft x-ray RIXS at 931~eV (the energy of the Cu $L_3$-edge resonance) has improved by over a factor of 10 from 1996 to 2008 to the current value of 130~meV at the Swiss Light Source \cite{Ghiringhelli2006, Ghiringhelli2012SAXES}. This has made it possible to access spin flip excitations in the cuprates \cite{Braicovich2010}. Figure \ref{Fig:Braicovich} plots RIXS measurements of the magnon dispersion of La$_2$CuO$_4$ taken from Ref.~\cite{Braicovich2010}, which are in excellent agreement with a spin wave fit to inelastic neutron scattering (INS) measurements \cite{Headings2010}. Shortly after the observation of magnons in La$_2$CuO$_4$, magnons were also observed in RIXS measurements of other square lattice antiferromagnetic insulating cuprates Sr$_2$CuO$_2$Cl$_2$ \cite{Guarise2010} and Bi$_2$Sr$_2$YCu$_2$O$_8$ \cite{Piazza2012}. Thus RIXS provides an alternative to INS for measuring the magnetic excitations in the cuprates.

INS is a very well established experimental technique and it can be performed with excellent energy resolution -- well below 1~meV, although the best practical energy resolution is usually limited by count rate considerations \cite{ShiraneBook}. Due to the weak interaction of the neutron with matter the INS cross section is relatively simple and well understood \cite{ShiraneBook}, which allows researchers to determine the magnetic dynamical structure factor, $S(\mathbi{Q},\omega)$ in absolute units \cite{Xu2013}. However, this weak interaction also means that neutrons travel several cm through matter before scattering and that large single crystals, several cm$^3$ in volume, are required for most INS experiments on cuprates.

In RIXS, on the other hand, the interactions are both stronger and more complicated and most theories for direct RIXS are based on operator treatments \cite{deGroot1998, Luo1993, Brink2006, Ament2009, Haverkort2010, Igarashi2012}. This strong interaction also means that the penetration depth of soft x-rays is of the order of 1000~\AA{} -- far shorter than that of thermal neutrons. This facilitates studies of small samples, indeed, hetrostructures based on 1 unit cell thick cuprate layers can even be measured \cite{Dean2012, Minola2012}. In point of fact, comparing the count rate, normalized to the probed volume, in state of the art RIXS \cite{Dean2012} and INS \cite{Headings2010} one finds that RIXS is $\sim10^{11}$ times more sensitive. This high sensitivity, even in small samples, is perhaps the key advantage of RIXS compared to INS, and it is this that has given Cu $L_3$-edge RIXS an important niche for measuring magnetic excitations, despite the relatively coarse energy resolution of the current RIXS spectrometers and the less well understood cross section.

\begin{figure}
\includegraphics[width=\linewidth]{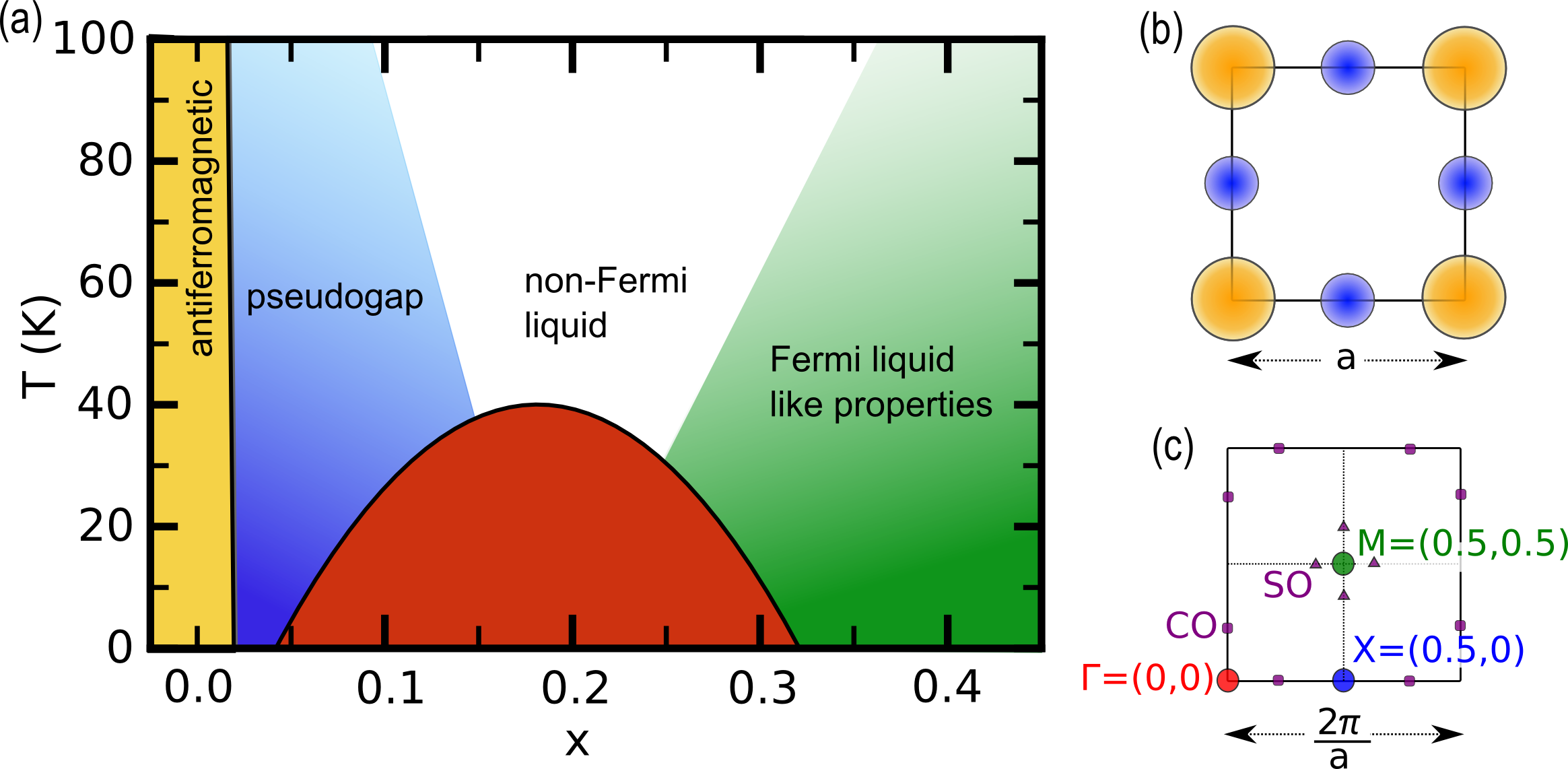} %
\caption{(a) The phase diagram for La$_{2-x}$Sr$_x$CuO$_4$ as a function of doping, $x$, showing antiferromagnetic, pseudogap, superconducting, non-Fermi liquid and Fermi liquid like phases. (b) The basic structural unit common to all HTS cuprates: a CuO$_2$ plaquette with Cu atoms shown in orange and oxygen atoms shown in blue. (c) The cuprate Brillouin zone with high symmetry points denoted by circles and labeled in reciprocal lattice units (r.~l.~u.). The charge order (CO) and spin order (SO) wavevectors present in 214-type cuprates such La$_{2-x}$Ba$_x$CuO$_4$ $x=1/8$ are shown by purple squares and purple triangles respectively. Throughout this review we will refer to $\mathbi{Q}$-vectors in terms of this Brillouin zone, often called the high temperature tetragonal structure.}
\label{Fig:cuprates}
\end{figure}
In terms of measuring magnetic excitations it is important to note that Raman scattering can also be used to measure magnetic excitations -- a field that is reviewed in,  for example, Refs.~\cite{Lemmens2003, Gozar2005}. In Raman scattering measurements on the cuprates, two magnon excitations appear most strongly (see for example Ref. \cite{Sugai2003}) through the spin exchange scattering mechanism \cite{Fleury1968}. Single magnon excitations are only very weakly allowed due to finite spin-orbit coupling in the valance band \cite{Fleury1968, Gozar2004}. In Raman scattering, however, the incident light carries negligible momentum so it can only probe the Brillouin zone center. Cu $L_3$-edge RIXS has a less strict limitation: it can, in principle, probe magnetic excitations out to $(0.5, 0)$, though it cannot reach the antiferromagnetic ordering wavevector $(0.5, 0.5)$. Cu $K$-edge RIXS has also been used to measure multimagnon excitations with a total summed spin of zero \cite{Harada2002, Hill2008, Ellis2010}, covering several Brillouin zones in reciprocal space. Cu $M$-edge \cite{Kuiper1998,Chiuzbaian2005} and O $K$-edge RIXS \cite{BisogniLCO2012, BisogniLSCO2012} can also detect multimagnon excitations although these have even more severe momentum restrictions than Cu $L$-edge RIXS.
\begin{figure*}
\includegraphics[width=0.7\linewidth]{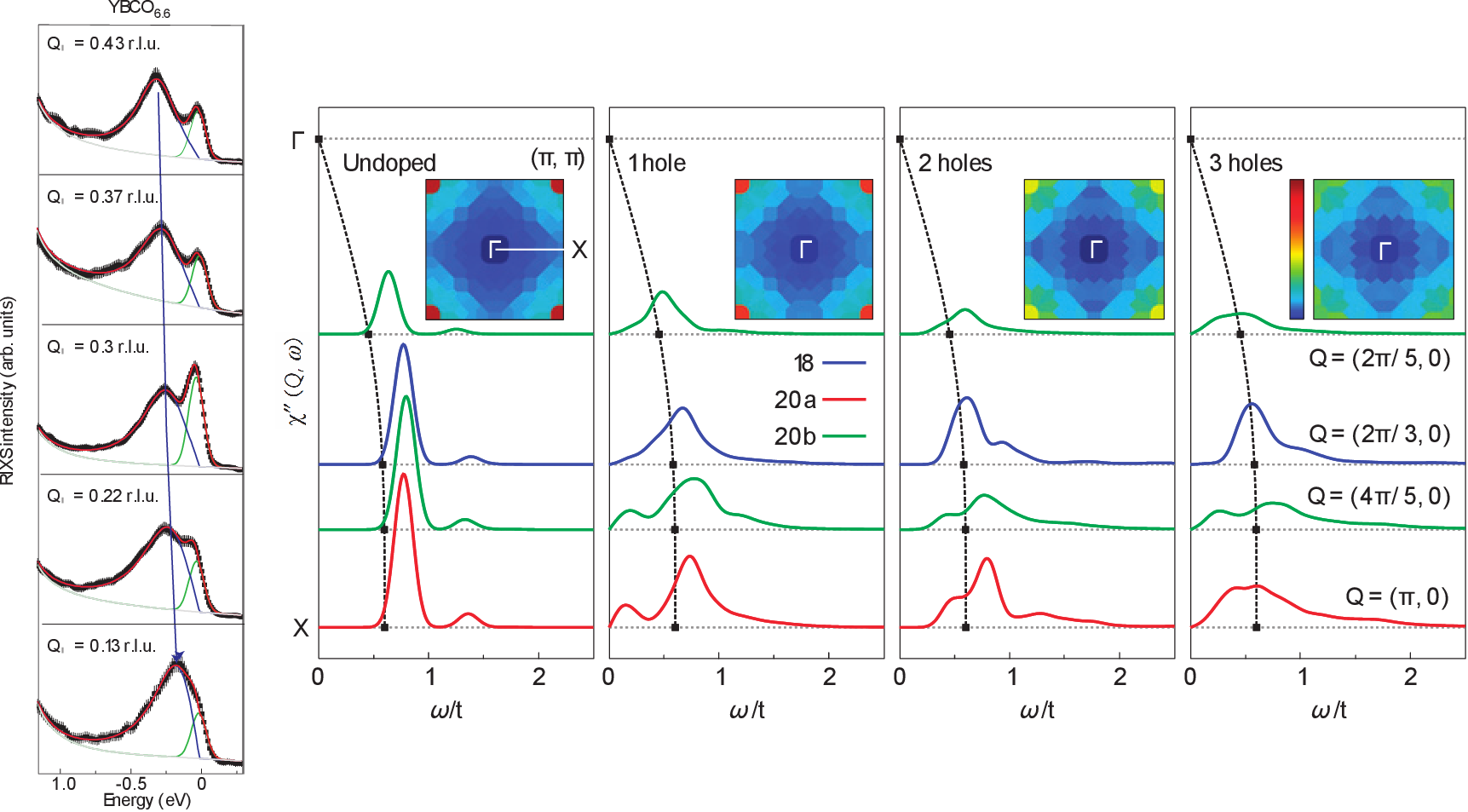} %
\caption{RIXS data and calculations from Ref.~\cite{LeTacon2011}. Left: RIXS measurements for YBa$_2$Cu$_3$O$_{6.6}$ along $(Q_{\parallel},0)$. Right: Cluster calculations of the imaginary part of the magnetic dynamical susceptibility $\chi^{\prime\prime}$ based on the $t-J$ model.}
\label{Fig:LeTacon}
\end{figure*}

\subsection{Cuprates\label{Sec:Cuprates}}
The  phase diagram of the cuprates is shown as a function of doping, $x$, in Fig.~\ref{Fig:cuprates}(a). Much of the physics of the cuprates, especially at low dopings, is dominated by a strong on-site Coulomb repulsion, $U$, between electrons, and many researchers believe that the three band Hubbard model, or simplified effective models based on this starting point, contains the physics required to describe the intrinsic properties of the normal state and the resulting HTS \cite{Dagotto1994}. The parent compounds of the cuprates such as La$_2$CuO$_4$ have one hole per Cu site and the strong U leads to insulating behavior with a charge excitation gap of 2~eV \cite{Kastner1998}. These localized holes, of predominantly Cu character \cite{Chen1992}, order antiferromagnetically below 325~K. The resulting spin dynamics is well described in terms of spin wave, or magnon, excitations \cite{Coldea2001, Headings2010} within the 2 dimensional spin $\frac{1}{2}$ Heisenberg model on the square lattice. Indeed, the Heisenberg model can be derived from the Hubbard model at zero doping (also often referred to as half filling) \cite{Dagotto1994}. At the other extreme, for $x \gtrsim 0.3$ the ground state and low energy electronic excitations have Landau Fermi liquid-like properties with resistivity that scales as $T^2$ and a fully connected Fermi surface \cite{Hussey2003}. The superconducting dome extends over $0.05 < x < 0.3$ where the superconducting state has a $d_{x^2-y^2}$ gap symmetry. Most of the complexity of the cuprates lies in the normal state from which HTS emerges.

 At a doping of $x=0.03$ antiferromagnetic order gives way to the pseudogap phase \cite{Timusk1999,Norman2005, Alloul2013, Yoshida2012,Norman2011, Norman2013}. This phase is associated with a partial reduction of the electronic density of states as the sample is cooled from high temperature through the pseudogap temperature. The electronic spectral function in this state has been extensively measured by ARPES \cite{Yoshida2012} which observes arcs of photoemission intensity, with strong spectral weight along the nodal ($(0,0)\rightarrow(0.5,0.5)$) direction and vanishing intensity along the antinodal ($(0,0)\rightarrow(0.5,0)$) direction. The origin of the pseudogap remains controversial, some popular interpretations are that it arises from preformed Cooper pairs that have not gained phase coherence \cite{Emery1995}, or that it comes from a competing ordering tendency such as charge/spin density waves \cite{Tranquada1995,Ghiringhelli2012} or loop currents \cite{Varma1997}. In many cuprates the presence of competing orders is well established \cite{Vojta2009} and the discussion has become focused on the relationship between competing order and HTS. Stripe order has been particularly well characterized in compounds such as La$_{1.875}$Ba$_{0.125}$CuO$_4$ where it manifests itself as charge order Bragg peaks at $(0.24,0)$  and spin order Bragg peaks at $(0.38,0.5)$ as shown on Fig.~\ref{Fig:cuprates}(c). As the doping level increases towards optimal doping, competing orders tend to disappear, as does the pseudogap, disappearing somewhere between $0.16 < x < 0.30$ depending largely on which probe one is using. This gives way to a strange metallic phase characterized by linear resistivity often called a non-Fermi liquid, which crosses over continuously into a metal with $T^2$  Fermi liquid resistivity at $x\approx 0.30$ \cite{Cooper2009}.

\begin{figure*}
\floatbox[{\capbeside\thisfloatsetup{capbesideposition={right,top},capbesidewidth=4cm}}]{figure}[\FBwidth]
{\caption{(a) RIXS spectra of optimally doped ($T_C = 92$~K) Bi$_2$Sr$_2$CaCu$_2$O$_{8+\delta}$ along the $(Q_{\parallel},0)$ symmetry direction (black dots). The spectra are offset for clarity. (b) A comparison of the peak in the RIXS dispersion (white $\square$) with calculations of $S(\mathbi{Q}, \omega)$ \cite{DeanBSCCO2013, James2012, James2013} based on the Yang-Rice-Zhang Anstaz \cite{Yang2006}. (c) The YRZ form for the electronic structure of Bi$_2$Sr$_2$CaCu$_2$O$_{8+\delta}$, based on fits to ARPES data in Ref.~\cite{Yang2008}. The intensity of color is proportional to the coherent quasi-particle weight and the gray surface depicts the Fermi energy. The arrow shows a typical magnetic scattering transition and all allowed magnetic transitions are summed over to form panel (b).  Copyright 2013 by The American Physical Society from Ref.~\cite{DeanBSCCO2013}.}\label{Fig:BSCCO}}
{\includegraphics[width=10cm]{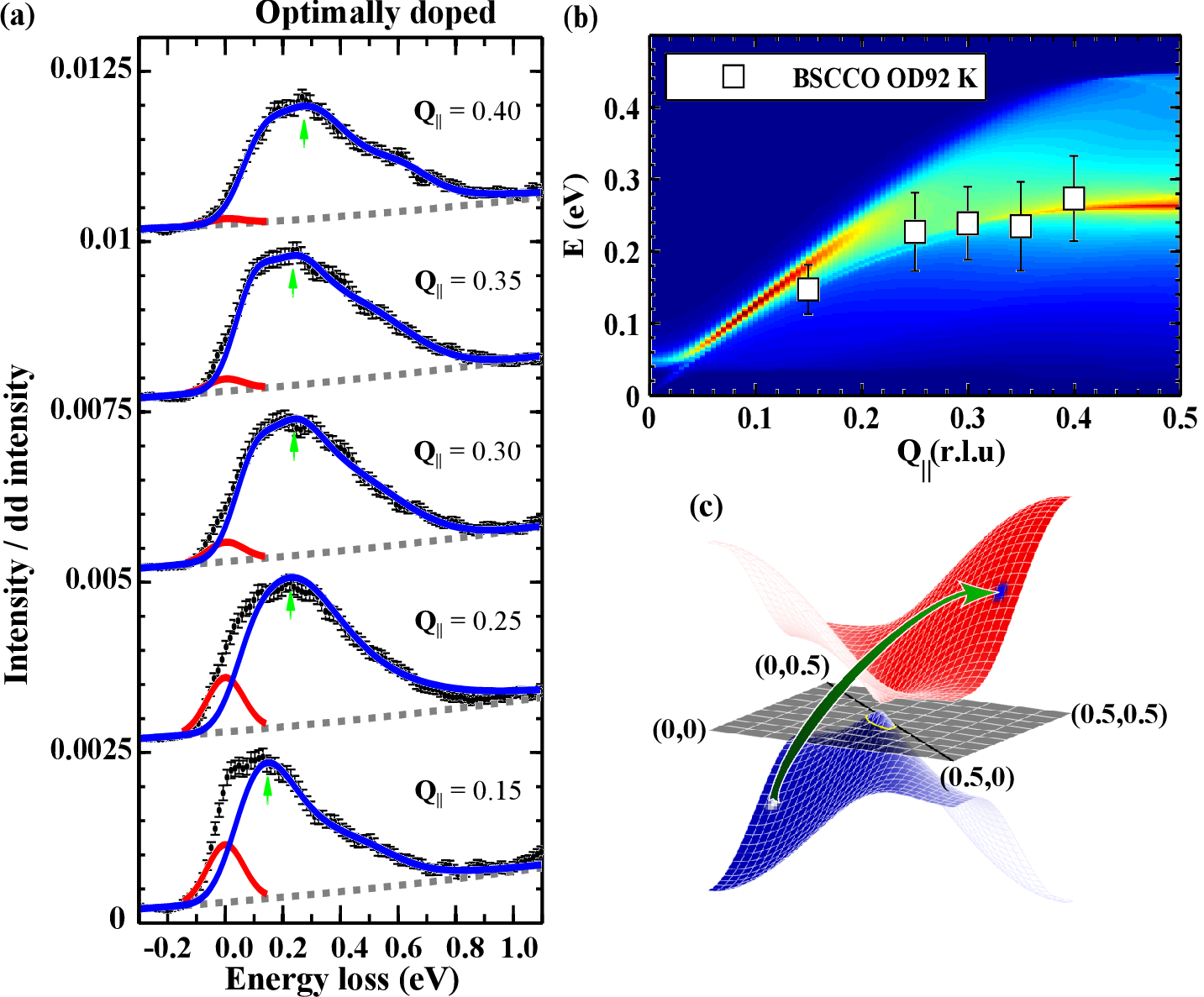}}
\end{figure*}

\section{Underdoped and optimally doped cuprates \label{Sec:Underdoped}}
As discussed in Section \ref{Sec:RIXS} most of the early Cu $L$-edge RIXS studies of the cuprates focused on undoped systems \cite{Braicovich2010, BraicovichPRB2010, Guarise2010, Piazza2012} where the excitations are understood in terms of spin wave or magnons \cite{Coldea2001}. Soon after magnons were measured in undoped cuprates, interest moved to also studying cuprates in the underdoped \cite{Braicovich2010,LeTacon2011,Dean2012,DeanLBCO2013} and optimally doped regions \cite{LeTacon2011, DeanBSCCO2013}. In this section, we first describe the experimental observations and their interpretation in hole doped cuprates and we then go on to their implications. At the end of this section we also address electron doped systems.

\subsection{Experimental observations}
Some of the first results on underdoped La$_{2-x}$Sr$_x$CuO$_4$ exhibited two dispersive features with energies below and above the magnon in La$_2$CuO$_4$ \cite{Braicovich2010}. These features were assigned to dynamical phase separation where the lower (higher) energy features come from regions of the CuO$_2$ planes with higher (lower) hole concentration. In light of subsequent work, it seems that this two peak structure is not a generic feature of the cuprates, rather it may be linked to structural disorder, which is known to be particularly strong in the 214 family of cuprates \cite{Alloul2009}. Studies of other cuprate samples with deliberately introduced disorder might further elucidate this issue. All subsequent studies of hole doped and electron doped cuprates report one peak in the mid infra-red region of the spectrum below 500~meV. These include measurements on the Yttrium and Neodymium-based cuprates \cite{LeTacon2011}, Bi$_2$Sr$_2$CaCu$_2$O$_{8+\delta}$ \cite{DeanBSCCO2013}, La$_{2-x}$Sr$_x$CuO$_4$ \cite{DeanLSCO2013}, La$_{2-x}$Ba$_x$CuO$_4$ \cite{DeanLBCO2013} and Nd$_{2-x}$Ce$_x$CuO$_4$ \cite{LeearXiv2013, IshiiESRF2012}. In general, underdoped and optimally doped cuprates exhibit qualitatively the same spectra. Example spectra for Yttrium based cuprates are shown in Fig.~\ref{Fig:LeTacon} \cite{LeTacon2011} and are very similar to subsequent measurements on Bi$_2$Sr$_2$CaCu$_2$O$_{8+\delta}$ plotted in Fig.~\ref{Fig:BSCCO}(a) \cite{DeanBSCCO2013}. A single peak is observed dispersing from a little over 100~meV around $(0.1, 0)$ to around 300~meV near to $(0.4, 0)$. This is very similar behavior to the magnon measured in La$_2$CuO$_4$ \cite{Braicovich2010,Dean2012} and for this reason, this peak is often referred to as a ``paramagnon'' where the ``para'' prefix denotes that the excitation arises from a state with short range rather than long range magnetic correlations. Consequently, the paramagnon excitation has a substantial linewidth, in contrast to the resolution-limited magnon in undoped cuprates.  At this point it is worth noting that once the cuprates are doped and become metallic, low energy charge excitations are possible. In the absence of a completely rigorous and generic understanding of the RIXS cross-section one might question whether the observed paramagnon peaks are indeed magnetic in nature. However, the smooth, continuous doping evolution of the intensity, energy and width of the peak, from the undoped values, along with the consistent behavior of the polarization dependence of the peak in several different cuprates make for a strong, if solely empirical, case that the peak is indeed magnetic in character \cite{LeTacon2011, DeanBSCCO2013, DeanLSCO2013, LeTacon2013, DeanLSCO2013}. More recently, well justified calculations of the RIXS intensity in doped cuprates have further strengthened this viewpoint \cite{Jia2013}. In addition $L$-edge RIXS experiments on other compounds with different electronic structures have also observed excitations that are completely consistent with the assertion that RIXS couples predominantly to magnetic excitations. These include Fe $L_3$-edge measurements of the pnictides \cite{Zhou2013} and Ir $L_3$-edge measurements of iridates \footnote{Spin wave excitations in insulating iridates have already been reported \cite{Kim2012, Kim2012_327, Yin2013} consistent with first principles calculations Ref.~\cite{Katukuri2012} and manuscripts describing similar observations for doped systems are in preparation by X.~Liu \emph{et al.} and J.~P.~Clancy \emph{et al.}}.

\subsection{Implications}
The observation of these paramagnons has several important implications. Most obviously, it demonstrates the existence of magnetic correlations up to optimal doping and beyond, a vital part of our characterization of the normal state from which superconductivity emerges. Prior to the RIXS studies discussed here, the existence of magnetic correlations at optimal doping was already known from extensive INS studies \cite{Fujita2012NS} (among other probes) and the existence and dispersion relations of excitations above 100~meV is shown particularly clearly in INS measurements using the MAPS spectrometer at the ISIS neutron source \cite{Vignolle2007, Lipscombe2007, Lipscombe2009}. RIXS has significantly extended the energy scale, $\mathbi{Q}$-range and doping range over which we have a good picture of the magnetic excitation spectrum. This data must be taken into account in theoretical models for magnetism in the cuprates and theoretical proposals for HTS based on pairing by exchange of magnetic fluctuations \cite{Eschrig2006, Scalapino2012}. For example, Ref.~\cite{LeTacon2011} analyzed measurements of various underdoped, optimally doped and slightly overdoped Yttrium-based cuprates using cluster exact diagonalization calculations of the $t-J$ Hamiltonian, as shown in Fig.~\ref{Fig:LeTacon}. Using the Eliashberg equations, the superconducting $T_c$ was calculated and found to be in agreement with the measured value within a factor of two. As is almost always the case in theories of HTS, the approximations made in such an approach are open to debate. It is also important to note that the discussion of these results assumed that the magnetic dispersion from $(0,0) \rightarrow (0.5,0) $ is equivalent to that from $(0.5,0.5) \rightarrow (0.5,0)$, which amounts to assuming the excitations are the same in the full structural Brillouin zone and the reduced antiferromagnetic Brillouin zone. This is well justified in antiferromagnetically ordered undoped cuprates, but it should be treated with caution in the non magnetically ordered doped cuprates. The damping of the magnetic excitations, in particular, is likely to be different as the Fermi arcs phenomenology is not symmetric on reflection about the $(0.5,0) \rightarrow (0,0.5)$ line. In more general terms, testing RIXS results against theories of magnetically mediated HTS is likely to be an important future trend.

The observation of paramagnons also helps in efforts to reconcile the electronic and magnetic properties of the cuprates within a single model. Of all the high-$T_c$ cuprate superconductors, Bi$_2$Sr$_2$CaCu$_2$O$_{8+\delta}$ is the most easily cleaved and ARPES \cite{Yoshida2012} and STS \cite{Fujita2012STS} have given us a highly accurate characterization of the electronic structure of this cuprate. Bi$_2$Sr$_2$CaCu$_2$O$_{8+\delta}$ is, however, challenging to grow as large single crystals and there is consequently scant information about the  high-energy magnetic response ($>100$~meV) in this cuprate \cite{Xu2009}, a gap that RIXS helped to fill \cite{DeanBSCCO2013}.

 Theoretical models of doped cuprates often take the approach of treating the magnetic and electronic properties of the cuprates separately: local magnetic moment approaches are typically used to interpret the spin response at low energies \cite{Xu2009} and these calculations provide a natural description of the ``hourglass'' dispersion near $(0.5,0.5)$ \cite{Vojta2009}, but it is difficult to connect these theories with the itinerant quasiparticles seen by ARPES \cite{Yoshida2012}. Few methods allow for calculations of the electronic \emph{and} magnetic properties using a single set of parameters.  The Yang-Rice-Zhang (YRZ) phenomenology for the electronic structure \cite{Yang2006} has the advantage that it provides a relatively simple explanation for Fermi arcs observed by ARPES and an easy way to implement parametrization for the electronic structure of the cuprates \cite{Rice2012}. In Refs.~\cite{James2012, James2013} a slave boson treatment of the $t-J$ Hamiltonian, consistent with the YRZ phenomenology, was used in conjunction with parameters determined from fits of the YRZ model to ARPES data on  Bi$_2$Sr$_2$CaCu$_2$O$_{8+\delta}$ \cite{Yang2011} to predict $S(\mathbi{Q},\omega)$. A comparison between the theoretical calculations and RIXS data is shown in Fig.~\ref{Fig:BSCCO} \cite{DeanBSCCO2013}. In view of the lack of adjustable parameters, the agreement between the measured paramagnon energy and the prediction is excellent. Further, such an approach also captures the weak doping dependence of the paramagnon. The $\mathbi{Q}$-dependence of the intensity and the width of paramagnon is currently challenging to compare in detail with the theory. There are several future directions that are likely to improve the agreement between theory and experiment. Incorporating the possible appearance of an ordered moment \cite{Eremin2013} in the YRZ description at low doping should improve the accuracy of the theory for $\mathbi{Q}\rightarrow (0,0)$, which is currently dominated by the large pseudogap. The incoherent part of the Green's function would also ideally be included, which is likely to broaden the paramagnon feature. Explicit calculations of the RIXS cross section, rather than simply $S(\mathbi{Q},\omega)$, would also be valuable as this would allow for detailed intensity comparisons.

\begin{figure*}
\centering
\includegraphics[width=0.85\linewidth]{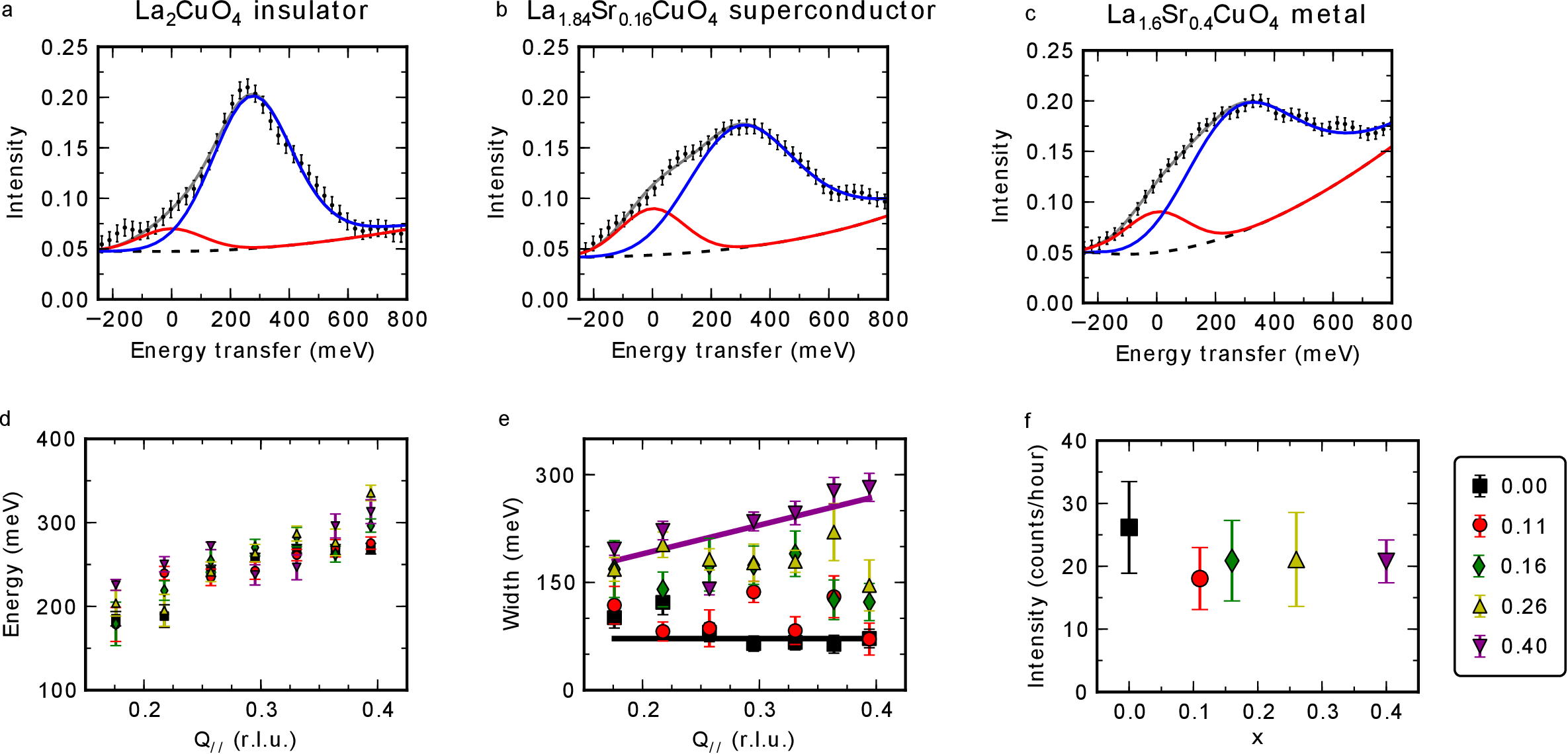} %
\caption{RIXS spectra at 25~K and $\mathbi{Q}=(0.33, 0)$ for La$_{2-x}$Sr$_x$CuO$_4$ at (a) $x=0$ (insulating) (b) $x=0.16$ (superconducting) (c) $x=0.40$ (metallic non-superconducting) samples. The filled black circles represent the data and the solid grey line shows the results of the fitting, which is the sum of an elastic line (red), magnetic scattering (blue), and the background (dashed black). The bottom row of panels plot the energy (d), width (e) and intensity (f) of the magnetic excitations, as determined by fitting. This figure is based on data and analysis from Ref.~\cite{DeanLSCO2013}.}
\label{Fig:LSCO}
\end{figure*}

\subsection{Electron doped cuprates}
Although the majority of cuprates are hole doped, a small number of cuprates can also be prepared in an electron doped form such as Nd$_{2-x}$Ce$_x$CuO$_4$ \cite{Armitage2010}. Given that these compounds also exhibit HTS, they form an interesting point of comparison. Although in the nearest neighbor single band Hubbard model the effects of electron and hole doping should be symmetric, once the oxygen and Cu states are included as separate bands, or once next nearest neighbor hopping terms are included this is no longer the case. Cu $L_3$-edge RIXS experiments on electron doped cuprates have only been performed very recently \cite{LeearXiv2013, Ishii2014}. The peak associated with the magnon in Nd$_{2-x}$Ce$_x$CuO$_4$ with $x=0.04$ is observed to harden by approximately 50\% in energy with $x=0.15$ electron doping \cite{LeearXiv2013}. Such a result is in many ways counterintuitive as doping might be thought of as causing a spin vacancy, reducing the number of Cu-Cu magnetic bonds, and reducing the energy required for a spin-flip excitation. However, this result is reproduced in cluster calculations of the Hubbard model and explained in terms of three-site exchange \cite{Jia2013}.

\section{Overdoped cuprates\label{Sec:Overdoped}}
After the insights gained in studies of underdoped and optimally doped cuprates, interest was naturally extended into the overdoped region. We will again divide this section into experimental observation and the implications, in which we also address the comparison between $L$-edge RIXS and probes sensitive to two magnon scattering.

\subsection{Experimental observations}
Characterizing the magnetic excitations in the overdoped cuprates is particularly interesting in light of the phase diagram in Fig.~\ref{Fig:cuprates}. While the disappearance of HTS in the underdoped regime is likely due to the proximity of the insulating state, the reason for the disappearance of HTS with overdoping is far less clear. Indeed, within the BCS model for conventional superconductivity an increase in the electronic density of states increases $T_c$, so the reduction in $T_c$ on the overdoped ($x > 0.16$) side of the cuprate phase diagram, may well be driven by a reduction in the strength of the pairing interaction. This motivated many INS experiments which showed that although magnetic excitations persist up to around optimal doping \cite{Bourges2000, Tranquada2004, Hayden2004, Vignolle2007, Xu2009}, they effectively disappear in the overdoped cuprates  \cite{Wakimoto2004, Wakimoto2007, Fujita2012NS}, and this was argued to cause the demise of HTS with overdoping \cite{Wakimoto2004, Fujita2012NS, Scalapino2012}. Such a scenario motivated RIXS studies of overdoped cuprates \cite{DeanLSCO2013, LeTacon2013}. Figure \ref{Fig:LSCO} shows the results of RIXS measurements on samples crossing the whole La$_{2-x}$Sr$_x$CuO$_4$ doping phase diagram (Fig. \ref{Fig:cuprates}) from $\mathbi{Q} = (0.15, 0) \rightarrow (0.40,0)$ \cite{DeanLSCO2013}. The spectra, and the fitting analysis, show that the paramagnon  persists across the whole doping phase diagram with comparable spectral weight and similar energies. The width on the other hand, increases continuously, consistent with the paramagnon being damped by the increasing electronic density of states. Similar paramagnon excitations were also observed Tl$_2$Ba$_2$CuO$_{6+\delta}$ \cite{LeTacon2013}.

\subsection{Implications}
The most important implication of these RIXS measurements is that the overdoped cuprates are not simple non-magnetic metals, but rather substantial magnetic correlations persist in the region of $\mathbi{Q}$ space probed in these experiments. This suggests that the high energy paramagnons measured by RIXS are unlikely to be a major factor in the pairing interactions, as these retain roughly constant energy and spectral weight as a function of $x$ while superconductivity disappears with overdoping. Only the lifetime of the paramagnon excitations changing appreciably. The change in $T_c$ is consequently most likely driven by other factors. These could include, among other things, the influence of the low-energy magnetic excitations which \emph{do} change dramatically in the overdoped cuprates \cite{Wakimoto2004, Wakimoto2007, Fujita2012NS}.

These results should also be considered in light of other probes of magnetic excitations. In this regard, it is vital to emphasize that RIXS has been used, predominantly, to measure the $(0,0) \rightarrow (0,0.5)$ symmetry line. Future RIXS studies will be important in clarifying the behavior along the other primary symmetry direction $(0,0) \rightarrow (0.3,0.3)$. Unfortunately, momentum conservation at the fixed incident energy dictated by the resonance prevents Cu $L_3$-edge RIXS from reaching the $(0.5,0.5)$ point, which is the antiferromagnetic ordering wavevector in undoped cuprates and the location of the highest intensity magnetic excitations. Taking the spin wave parameterization of La$_2$CuO$_4$ determined by fitting the INS data to the Heisenberg model \cite{Coldea2001}, we find that 27\% of the magnetic spectral intensity in $\chi^{\prime\prime}(\mathbi{Q},\omega)$ lies in a circle with radius 0.4 r.l.u.\ about $(0,0)$ \cite{DeanLSCO2013} i.e.\ the region easily accessible to RIXS. In contrast, INS experiments on the doped cuprates focus around $(0.5,0.5)$ while the intensity of the excitations along the $(0,0) \rightarrow (0,0.5)$ symmetry direction is typically below the signal-to-noise ratio of state-of-the-art INS measurements. So to date, there have not been any direct comparisons of RIXS and INS on doped cuprates at the same \mathbi{Q}.

Taken together, RIXS and INS data imply that doping does not uniformly attenuate the intensity of the magnetic excitations. Rather, the high-energy excitations around $(0.18\rightarrow0.40, 0)$ remain relatively constant in integrated intensity while the absolute intensity of the lower energy excitations around $(0.5,0.5)$ are strongly reduced. The energy dispersion of the magnetic excitations along $(0.18\rightarrow0.40, 0)$ also remains Heisenberg-like and roughly constant as as function of doping, while the dispersion near $(0.5,0.5)$, changes dramatically to form the ``hour-glass'' feature \cite{Arai1999,Bourges2000, Tranquada2004, Hayden2004, Vignolle2007, Xu2009}.

Although, the persistence of the paramagnon as seen in RIXS was surprising in the context of the previous INS studies, it is seen in calculations based on the single band Hubbard model. For example,  dynamical quantum Monte Carlo cluster calculations of $S(\mathbi{Q},\omega)$ show a paramagnon that has minimal change in energy and spectral intensity around $(0.5,0)$, while the excitations around $(0.5,0.5)$ are strongly attenuated \cite{scalapino2007}. Exact diagonalization cluster calculations of the RIXS intensity show a similar phenomenology and also exhibit excellent agreement with RIXS experiments \cite{Jia2013}. This further strengthens the already convincing case that the paramagnon peak in the cuprates is indeed magnetic in nature \cite{Jia2013}. On the other hand, Chen and Sushkov argue that the $t-J$ model, which can be derived from the Hubbard model by projecting out doubly occupied sites, shows an appreciable softening of zone boundary magnetic excitations around $(0.5,0)$ at optimal doping and that inconsistent with the RIXS observations \cite{Chen2013}. This was interpreted as arising from a breakdown of the Zhang-Rice singlet approximation \cite{Zhang1988}  outside of the heavily underdoped regime \cite{Chen2013}. Work is in progress to extend $t-J$ model calculations to higher hole dopings and electron doping \cite{LeeMarchMeeting}.

The Cu $L_3$-edge RIXS results can also be compared to other probes which couple to two magnon processes including Raman scattering and $K$-edge (and in principle $M$-edge) RIXS. Raman scattering is sensitive to the two magnon density of states at $\mathbi{q}\approx0$, which in a simple Heisenberg antiferromagnet with only nearest neighbor exchange interactions, $J$, will lead to a peak at $4J$ \cite{Lemmens2003}.

 In the undoped cuprates such as La$_2$CuO$_4$ the two magnon Raman signal consists of an asymmetric peak around 400~meV \cite{Lyons1988}. Due to fact that Raman scattering creates magnons in close proximity to each other in real space it is often argued that magnon-magnon interactions reduce the energy of the two magnon peak \cite{Elliott1968} and, when including this effect along with longer range ring exchange interactions, the energy of the two magnon peak can be reproduced based on spin wave theory \cite{Katanin2003}. However, although the energy of the peak in La$_2$CuO$_4$ is well understood, the width of the peak is yet to be reproduced without invoking more complicated effects such a triple resonance effect in the Raman cross section or additional damping of the magnon due to coupling to phonons or electronic excitations \cite{Chubukov1995}. Several studies have extended these observations to the doped cuprates typically using the same incident laser energy. In this case, the intensity and energy of the two magnon peak drops strongly with doping \cite{Cooper1993, Blumberg1994, Rubhausen1997, Sugai2003, Muschler2010}. 
 
Recent studies of HgBa$_2$CuO$_{4+\delta}$, however, have argued that ellipsometry measurements imply that the appropriate incident laser energy to observe the two magnon peak changes significantly with doping \cite{Li2013}. By performing measurements at different incident laser energies the magnon peak in HgBa$_2$CuO$_{4+\delta}$ has been shown to persist with appreciable intensity and energy into the overdoped regime $x=0.19$ \cite{Li2013}. In the electron doped system Nd$_{2-x}$Ce$_x$CuO$_4$ the two magnon peak has been observed from $x=0$ to $x=0.10$ without evidence for strong softening of the magnetic excitations with doping \cite{Onose2004} . Reconciling the hole and electron doping results seen by two magnon Raman scattering with $L$-edge RIXS remains an important issue that deserves more attention in future studies. 

Two magnon excitations are also the dominant magnetic excitations in $K$-edge RIXS in which the lack of spin-orbit coupling of the core hole means that the single magnon process is forbidden.  Theoretical studies then explain the coupling to the magnetic excitations in terms of the perturbing effect of the core hole \cite{Brink2007, Ament2007, Nagao2007, Forte2008}. O $K$-edge RIXS studies show a similar phenomenology to the Raman results in Ref.~\cite{Li2013}, in which a two magnon peak in La$_{2-x}$Sr$_x$CuO$_4$ was observed with similar energy and roughly comparable intensity from $x=0$ to $x = 0.22$. In this case it was also necessary to change the incident excitation energy in order to observe the peak \cite{BisogniLCO2012,BisogniLSCO2012}. Finally, two magnon excitations have also been observed at the Cu $K$-edge \cite{Harada2002, Hill2008, Ellis2010} . $K$-edge measurements tracking the peak evolution with doping find that it disappears in La$_{2-x}$Sr$_x$CuO$_4$ at $x=0.07$, although in this case broad peaks can be difficult to observe due to the strong elastic scattering \cite{Hill2008, Ellis2010}. $K$-edge RIXS measurements as a function of incident x-ray energy might be important for resolving this issue.  In the future, M-edge RIXS might also become a viable probe of the doping dependence of magnetic excitations in the cuprates, but to date magnetic excitations have only been observed in concert with $dd$ orbital transitions at the Ni and Cu $M$-edge in undoped cuprates and nickelates \cite{Kuiper1998,Chiuzbaian2005}.

\section{Future trends}
\begin{figure*}
\floatbox[{\capbeside\thisfloatsetup{capbesideposition={right,center},capbesidewidth=4cm}}]{figure}[\FBwidth]
{\caption{The schematic for the layout of the Soft Inelastic X-ray (SIX) beamline at the National Syncrotron Light Source II. Figure courtesy of J. Dvorak, I. Jarrige and S. Pjerov.}\label{Fig:SIX}}
{\includegraphics[width=12cm]{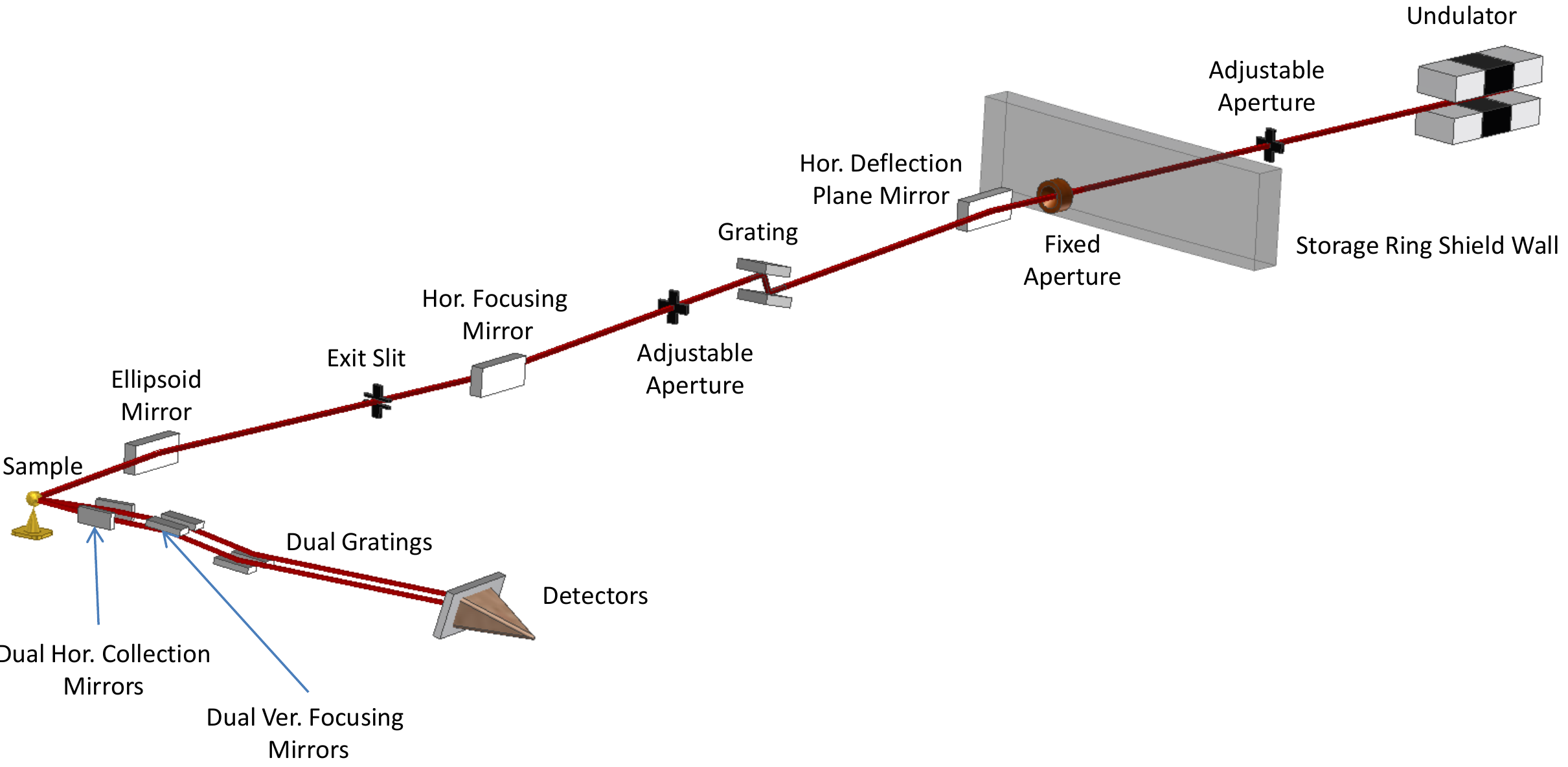}}
\end{figure*}

The insights gained using soft x-ray RIXS has generated considerable enthusiasm for expanding and improving RIXS instrumentation. One such project is the CENTURION spectrometer at the Soft Inelastic X-ray (SIX) beamline at the National Synchrotron Light Source II (NSLS-II), which is designed to achieve 14~meV energy resolution at the Cu $L_3$-edge \cite{SIX} almost a factor 10 higher than the current state-of-the-art, which at the time of writing is 130~meV at the ADRESS beamline at the Swiss Light Source \cite{Ghiringhelli2006}. Elsewhere, projects such as ERIXS at the European Synchrotron Radiation Facility \cite{ERIXS}, I21 at the Diamond Light Source \cite{I21},  VERITAS at MAX-IV \cite{VERITAS} and AGM-AGS at the Taiwan Photon Source \cite{Huang2014, Lai2014} are also aiming for very high energy resolution. Several of these projects will offer additional advantages such as the ability to continuously rotate the spectrometer to access different $|\mathbi{Q}|$, something that typically requires breaking vacuum at existing spectrometers, and the ability to analyze the scattered x-ray polarization \cite{Ghiringhelli2013CuL3rev}. Figure \ref{Fig:SIX} shows the schematic layout for the SIX beamline at the NSLS-II. X-rays are produced by an elliptically polarized undulator, and are dispersed onto the exit slit, the opening of which defines the incident x-ray bandwidth. Two mirrors focus the beam in the horizontal and vertical directions to obtain a spot size of 2.5 (horizontal) $\times$ 0.26 (vertical) $\mathrm{\mu m}^2$ at the sample. The spectrometer features two horizontal mirrors which collect the scattered x-rays and a grating that disperses the x-rays onto the CCD (charged coupled device), which collects a RIXS spectrum as a function of energy loss at a particular $\mathbi{Q}$ and incident energy. In order to achieve the planned 14~meV resolution, several key optical components, including the pre-mirror and the incident and spectrometer gratings, require state-of-the-art slope errors down to 0.05~$\mathrm{\mu}$rad and a very long (105~m) beamline.

These improvements in the next generation of soft x-ray RIXS spectrometers are likely to bring numerous new insights into HTS. Some possibilities for novel experiments are discussed below including experiments on stripe ordered cuprates and cuprate-containing heterostructures, as well as the possibility of measuring phonon excitations. This list is not meant to be comprehensive, for example we do not address the possibilities of observing the superconducting gap \cite{Marra2013} and the likely expansion of soft RIXS studies to measuring magnetic excitations in non-cuprate materials such as the manganites.

\subsection{Stripe correlations}
\begin{figure}
\centering
  \includegraphics[width=0.8\linewidth]{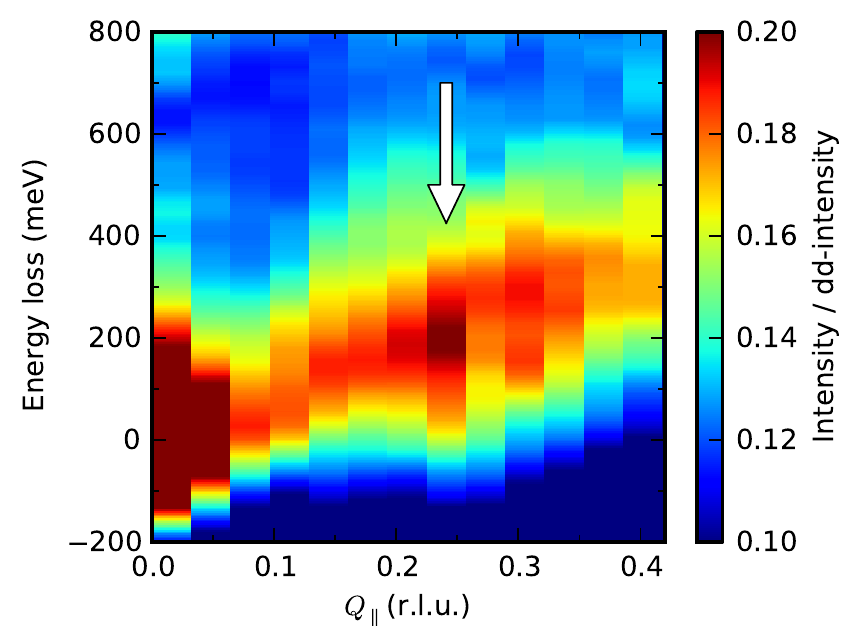} %
   \caption{RIXS measurements La$_{1.875}$Ba$_{0.125}$CuO$_4$ showing the dispersion of the magnetic excitations along $(Q_{\parallel},0)$ \cite{DeanLBCO2013}. The white arrow marks the charge order wavevector at $\Q_{\parallel}=0.24$ at which there is a possible renormalization of magnetic excitation spectrum by the charge ordering -- something that it is important to test for in future experiments. Copyright 2013 The American Physical Society \cite{DeanLBCO2013}.}
   \label{Fig:LBCO}
\end{figure}

As discussed in Section \ref{Sec:Cuprates}, underdoped cuprates show a tendency towards density-wave ordering, with resonant soft x-ray studies playing an important role in characterizing the nature of the order \cite{Vojta2009, FinkRepProgPhys2011, Seibold2012}. In La$_{2-x}$Ba$_x$CuO$_4$ in particular, charge and spin ordering is well established \cite{Abbamonte2005, Wilkins2011, Fujita2012stripes, Thampy2013} and there have already been extensive INS studies measuring how the magnetic excitations are renormalized by static stripe ordering around the magnetic ordering wavevector $(0.38, 0.5)$ \cite{Tranquada2004}. Far less is known, however, about the behavior around the charge-stripe ordering wavevector around $(0.24, 0)$ because here the intensity of the magnetic excitations are typically below the signal-to-noise level of inelastic neutron scattering experiments. This provides an excellent opportunity for Cu $L_3$-edge RIXS. Figure \ref{Fig:LBCO} plots initial measurements of the paramagnon excitation as it disperses through the stripe ordering wavevector \cite{DeanLBCO2013}. These results indicate that the magnetic excitation spectrum might be altered around the charge ordering wavevector, but the resolution in these experiments (260~meV full width at half maximum) was not good enough to disentangle possible changes in the magnetic excitations from the additional elastic intensity from the static stripe. Future higher resolution measurements will be vital in order to resolve this mystery. Such measurements can also, in principle, determine the distribution of anisotropic holes within the CuO$_2$ plane \cite{Seibold2006,Seibold2012} and, in particular,  distinguish bond and site centered stripes \cite{Tworzyd1999}. There is also an opportunity to identify excitations associated with the one-dimensional nature of the magnetic stripe itself. We note that such features have already been observed in the stripe-ordered La$_{\frac{5}{3}}$Sr$_{\frac{1}{3}}$NiO$_4$ using INS \cite{Boothroyd2003}.

\subsection{Heterostructures}
\begin{figure}
\centering
  \includegraphics[width=0.9\linewidth]{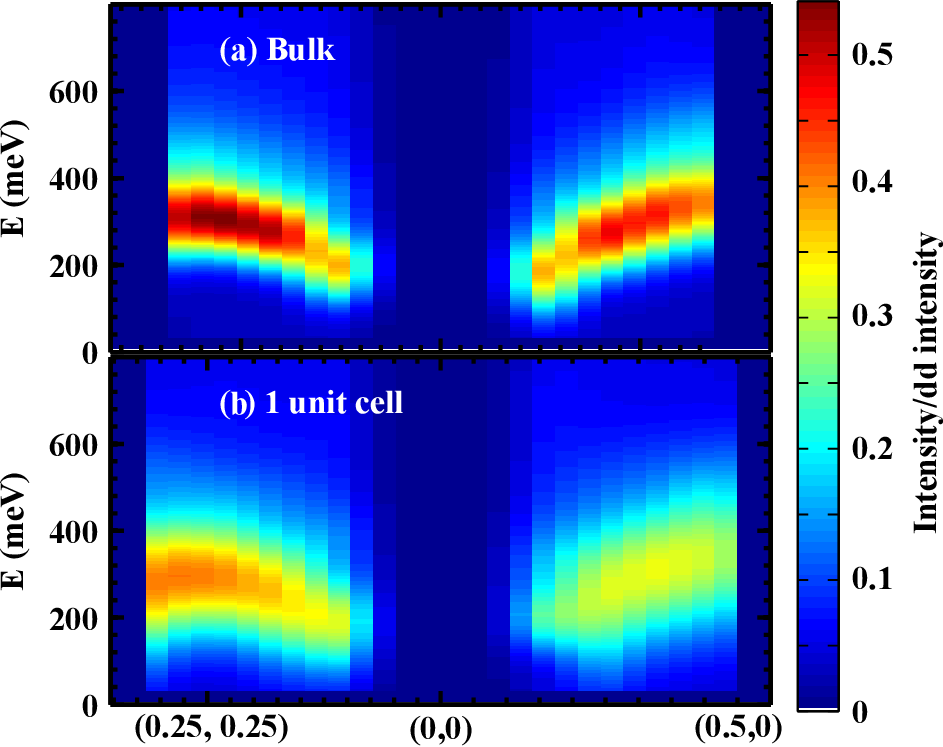}
   \caption{RIXS measurements of (a) a bulk-like 53~nm film of La$_2$CuO$_4$ and (b) a hetrostructure consisting  of 25 repeats of a LaAlO$_3$/La$_2$CuO$_4$ bilayer where each element in 1 unit cell thick. Figure adapted from Ref.~\cite{Dean2012}.
   }
   \label{Fig:hetrostructures}
\end{figure}

Advances in thin film deposition techniques via techniques such as molecular beam epitaxy and pulsed laser deposition now allow the synthesis of transition metal oxide layers with atomic precision. This provides an opportunity for further altering the cuprates, optimizing their existing emergent behaviors and even engineering completely new states \cite{Hwang2012}. In hetrostructures the properties of the cuprates can be tuned, not only by chemical doping, but also by charge transfer \cite{Gozar2008}, confinement within nanostructures \cite{Dean2012, Minola2012}, strain \cite{Bozovic2002}, orbital reconstructions \cite{Chakhalian2007}, exchange bias, and polar charge accumulation.  The additional complexity of these heterostructures calls for sensitive spectroscopic probes in order to characterize their properties in detail. Although we have several well established tools for measuring the electronic properties and \emph{static} magnetic order in heterostructures, their \emph{dynamic} magnetic correlations are not amenable to INS due to the impractically small sample volumes. To date, RIXS is the only technique that has been demonstrated to be capable of measuring the dispersion of magnetic excitations in transition metal oxide heterostructures. It is important to note that spin polarized electron energy loss spectroscopy can also measure collective spin excitations in thin films and heterostructures \cite{Vollmer2003}. Current setups use electrons of a few eV in energy, and they are therefore sensitive to the magnetic state at the surface (rather than the whole volume) of a film. Current research has focused on exotic surface phenomena in ferromagnetic metals such as iron and cobalt \cite{Zakeri2013}. It will be interesting to see whether this technique can provide insights into oxide systems in the future. In terms of RIXS, Fig.~\ref{Fig:hetrostructures} shows a measurement comparing the magnetic dispersions in bulk La$_2$CuO$_4$ with that of isolated one-unit-cell thick layers \cite{Dean2012}. This was one of the first studies to measure the dispersion of the magnetic excitations within a transition metal oxide heterostructure \cite{Dean2012, Minola2012}. RIXS experiments to date have only observed the relatively small changes induced in the CuO$_2$ in-plane exchange interactions of existing films. Perturbing the cuprates more dramatically has more potential for introducing novel effects, however, such samples are likely to be more difficult to synthesize. Nonetheless, the considerable potential of thin film synthesis in realizing novel new materials, and the anticipated progress in deposition techniques \cite{TsymbalBook},  will doubtless motivate more RIXS studies on hetrostructures in the near future.

\subsection{Phonons}
To date the vast majority of RIXS studies on the cuprates have focused on magnetic, charge and orbital excitations. This is largely due to energy resolution limitations, because the maximum phonon energy in the cuprates is $\sim90$~meV, well below the best energy resolution available at the Cu $L$-edge and only a little higher than the best energy resolution at the O $K$-edge. (Cu $K$- and $M$-edge work often has superior resolution, but more severe elastic line contributions to the spectra.) This is unfortunate because electron-phonon (e-ph) coupling has also been implicated as possibly contributing to HTS; either as a sole cause \cite{Lanzara2001} or working in concert with magnetic excitations \cite{Johnston2010}. RIXS, being bulk sensitive, element resolved and $\mathbi{Q}$ resolved, has the potential to provide additional information on top of what can be learned from non-resonant inelastic x-ray and neutron scattering, point contact spectroscopy and ARPES. Indeed, recent theoretical work suggests that one can extract the momentum-resolved e-ph coupling from RIXS spectra through the relative intensity scaling between the one, two and three phonon, etc.\ overtone peaks \cite{AmentPhonons2011}. Motivated by this potential, two preliminary RIXS experiments have been performed on cuprates including Ca$_{2+5x}$Y$_{2-5x}$Cu$_5$O$_{10}$ \cite{LeePRL2013} at the O $K$-edge and CuO at the Cu $K$-edge \cite{Yavas2010}, both of which provide evidence for multiple phonon overtones. Of course, current experiments combine the challenge of interpreting complex e-ph coupling phenomena with difficulties in interpreting the RIXS cross section for phonons, which remains only partially understood. In this regard, a RIXS measurement of a material with a well understood electron-phonon coupling will be vital for making progress in this emerging area.

\section{Concluding remarks}
We hope that this review has conveyed that RIXS has already made significant contributions to our understanding of magnetism and its relationship with HTS in the cuprates. Outside the scope of this review, RIXS studies have been important in determining the electronic properties of the cuprates, a field that has been reviewed in \cite{Ishii2013, Ament2011,Kotani2001}, and $L$-edge RIXS studies of non-cuprate materials such as iridates and pnictides have also been highly instructive. It has been a time of rapid progress for soft x-ray RIXS, fueled by improvements in energy resolution, and this pace of change looks set to continue in the coming years with the advert of multiple high-resolution RIXS beamlines. It promises to be an exciting time in the field.

\section{Acknowledgements}
The author would like to thank all those who he has collaborated with in RIXS studies in recent years especially his fellow Brookhaven scientists John Hill and Ivan Bo\v{z}ovi\'{c}. Special mention is also due to Lucio Braicovich, Valentina Bisogni, Jeroen van den Brink, Nick Brookes, Sorin Chiuzb\u{a}ian, Greta Dellea, Joe Dvorak, Giacomo Ghiringhelli, Henri Alloul, Mathieu Le Tacon, Yuan Li, Andrew James, Ignace Jarrige, Robert Konik, Thorsten Schmitt,  Ross Springell,  Andrew Walters and Kejin Zhou. This work is supported by the Center for Emergent Superconductivity, an Energy Frontier Research Center funded by the U.S. DOE, Office of Basic Energy Sciences and by the Office of Basic Energy Sciences, Division of Materials Science and Engineering, U.S.\ Department of Energy under Award No.\ DEAC02- 98CH10886. The use of the ADRESS beamline at the Swiss Light Source and the ID08 beamline at the European Synchrotron Radiation Facility is gratefully acknowledged.

\end{document}